\documentclass{PoS}
\usepackage{amsmath}

\title{Chiral methods at the electroweak scale}

\ShortTitle{Chiral methods at the electroweak scale}

\author{\speaker{Oscar Cat\`a}\thanks{This work was performed in the context of the ERC Advanced Grant project FLAVOUR (267104) and was supported in part by the DFG cluster of excellence EXC 153 'Origin and Structure of the Universe' and the Munich Institute for Astro- and Particle Physics (MIAPP).}\\
        Ludwig-Maximilians-Universit\"at M\"unchen, Fakult\"at f\"ur Physik,\\
Arnold Sommerfeld Center for Theoretical Physics, D80333 M\"unchen, Germany\\
        E-mail: \email{oscar.cata@physik.uni-muenchen.de}}

\abstract{I review the main features of the effective field theory (EFT) behind scenarios of dynamical electroweak symmetry breaking, placing particular emphasis on the systematics and the parallels that can be drawn with Chiral Perturbation Theory. The notion of chiral dimensions will be introduced and shown to be the right tool to describe nonlinear expansions. I will also discuss why such an EFT is of interest in phenomenological studies at the LHC. The most important aspect is that the EFT is engineered to recover the Standard Model in a particular limit, and therefore provides a general framework to test the Higgs hypothesis. Additionally, I will argue that the $\kappa$ formalism used currently by experimental collaborations to study Higgs couplings at the LHC can actually be embedded into this EFT. This not only gives the $\kappa$ parametrization a solid QFT foundation but also shows the way to improve it systematically, and in particular how to upgrade analyses on Higgs processes from the level of rates to the level of distributions.}

\FullConference{The 8th International Workshop on Chiral Dynamics, CD2015 \\
		29 June 2015 - 03 July 2015\\
		Pisa,Italy}

\begin{document}

%%%%%%%%%%%%%%%%%%%%%%%%%%%%%%%%%%%%%%%%%%%%%%%%%%%%%%%%%%%%%%%%%%%%%%%%%%%%%

\section{Introduction}

The idea that the pattern of electroweak symmetry breaking (EWSB) could be triggered by strong dynamics, in a way similar to what happens in QCD, is about to turn 40. The flagship for models of dynamical EWSB was technicolor~\cite{Weinberg:1979bn}, a scaled-up version of QCD from the GeV to the TeV scale. Technicolor was proposed as an alternative to the canonical Standard Model (SM) Higgs scenario and has been always associated with Higgsless theories. With the discovery of a light scalar its simplest versions are definitely ruled out, but that does not extend to the more general concept of dynamical EWSB.

In particular, dynamical EWSB scenarios can be easily extended to accommodate light scalars. Ideas in this direction were proposed already in the mid 80's with models of vacuum misalignment~\cite{Kaplan:1983fs}. In those models the Higgs is part of a Goldstone multiplet coming from an enlarged symmetry that gets broken at some scale $f$. The regular Goldstones become the longitudinal modes of the SM gauge fields, while the Higgs acquires a mass through loop processes, thus being naturally light. 

An explanation for the lightness of the Higgs is actually missing in the Standard Model and is one of the main motivations to search for new physics. Naive expectations set this new-physics scale at the TeV.  However, with the present accuracy on Higgs couplings (roughly at the $10\%$ level) the nature of both the Higgs and the interactions behind EWSB is something that is far from being settled. Both for weakly-coupled and strongly-coupled extensions there exist models which can satisfactorily explain the existence of a light scalar in a natural way, i.e., through symmetries (most notably supersymmetry for weakly-coupled scenarios and shift symmetry for strongly-coupled scenarios). However, it is also fair to say that the simplest versions of those models sooner or later run into other sorts of difficulties and presently no model stands out as a compelling candidate for new physics. 
 
A more conservative approach is provided by effective field theories. EFTs have a more modest scope and do not provide dynamical explanations: they are a model-independent tool to parametrize physics at a certain energy scale $\mu$. Provided that the particle content and symmetries relevant at energies of ${\cal{O}}(\mu)$ are known and that a mass gap up until a cutoff $\Lambda$ exists, one can build a meaningful expansion in $\mu/\Lambda$. EFTs are therefore a tool for phenomenological exploration in the search for new physics. As soon as new physics is detected, they cease to be valid and have to be upgraded. 

For weakly-coupled scenarios the corresponding EFT has been known for a while~\cite{Buchmuller:1985jz} and in recent years brought to a firm footing~\cite{Grzadkowski:2010es}. For strongly-coupled scenarios the first attempts were done long ago~\cite{Appelquist:1980vg} and later upgraded to include a Higgs-like scalar~\cite{Feruglio:1992wf} but a systematic treatment was lacking until recently~\cite{Buchalla:2012qq,Buchalla:2013rka}. In this paper I will review these recent attempts at a systematic EFT for dynamical EWSB. As the title of the talk indicates, most of the techniques have a counterpart in the more familiar theory of pions, ChPT. I will however show that there is an interesting cross-feeding between the two EFTs, and in particular that the richer dynamical framework of the electroweak interactions helps illuminate certain systematic aspects of ChPT in a different light.

The main message that I will try to convey is that, given the projected accuracy that will be reached at the LHC, the EFT presented here is the right tool to search for new physics during the LHC running time. The nonlinear EFT that I will discuss in this paper is not meant as an EFT of technicolor models, but rather as an EFT of vacuum misalignment dynamics. Interestingly, this makes this EFT a generalization of the Standard Model, i.e., the Standard Model is recovered for a specific choice of the parameters. The bigger parameter space spanned by the nonlinear realization of EWSB thus provides a framework in which the Higgs hypothesis can be tested in a model-independent way.   
      
%%%%%%%%%%%%%%%%%%%%%%%%%%%%%%%%%%%%%%%%%%%%%%%%%%%%%%%%%%%%%%%%%%%%%%%%%%%%%

\section{The Standard Model as a nonlinear EFT}

Nonlinear and linear EFTs are different in their systematics. By nonlinear EFTs I will be referring to theories with nonrenormalizable interactions and nondecoupling new dynamics. The distinction between them therefore goes well beyond the representation used for the fields. 

As an example, consider the Standard Model with the Higgs doublet $\phi$ and its charge-conjugate $\tilde{\phi}$ rewritten in polar coordinates through
\begin{align}
\phi=\frac{v+h}{\sqrt{2}}U\left(\begin{array}{c}0\\1\end{array}\right);\qquad \tilde{\phi}=\frac{v+h}{\sqrt{2}}U\left(\begin{array}{c}1\\0\end{array}\right)
\end{align} 
where $U$ is a $2\times 2$ matrix collecting the Goldstone modes. With this field redefinition one can express the Standard Model as
\begin{align}\label{SM}
{\cal L}_{SM}&=-\frac{1}{2}\langle W_{\mu\nu}W^{\mu\nu}\rangle 
-\frac{1}{4} B_{\mu\nu}B^{\mu\nu}+i\bar f_j \!\not\!\! Df_j+\frac{1}{2}\partial_{\mu}h\partial^{\mu}h\nonumber\\
&+\frac{v^2}{4}\ \langle D_\mu U D^\mu U^{\dagger}\rangle \left(1+\frac{h}{v}\right)^2-v\Big[{\bar{f}}_iy_{ij}UP_{\pm}f_j+{\mathrm{h.c.}}\Big]-\frac{\lambda}{4}(h^2-v^2)^2
\end{align}
The polar parametrization makes it manifest that the Standard Model, besides a local $SU(2)_L\times U(1)_Y$ symmetry, is also invariant under a global $SU(2)_R$ symmetry. The pattern of spontaneous symmetry breaking is therefore $SU(2)_L\times U(1)_Y\times SU(2)_R\to SU(2)_V\times U(1)_{{\mathrm{em}}}$, where the residual (global) $SU(2)_V$ is the so-called custodial symmetry. The biggest advantage of this field redefinition is that the radial (Higgs) and angular excitations (Goldstone modes) of the vacuum manifold are explicitly separated in a gauge-invariant way. The SM can therefore be expressed superficially as a nonlinear theory, but it is obvious that the dynamics has not changed. The theory is still renormalizable, new physics are decoupled and can be introduced as an expansion in canonical dimensions. In particular, one could consider the set of NLO (dimension-six) operators in polar notation. This would definitely obscure the renormalizable properties of the theory, but the Higgs and the Goldstone modes would still belong to a doublet under $SU(2)_L$. 

The theory becomes {\emph{dynamically}} nonlinear, i.e., nonrenormalizable, once the Higgs and the Goldstones do not belong to the same multiplet. This can easily be done with polar coordinates and amounts to generalize the Higgs couplings to arbitrary coefficients. In other words, the Higgs is demoted to an electroweak singlet. Intuitively, one can think of it as a two-step procedure: starting from Eq.~(\ref{SM}) one integrates out the SM Higgs mode and then reinstates a scalar particle with the most general couplings to gauge fields and fermions. In an abuse of language we will call this scalar the Higgs and denote it by $h$, though it should be clear that it does not necessarily correspond to the Standard Model one.  

The result for this more general theory reads
\begin{align}\label{LO}
{\cal L}_{LO}&=-\frac{1}{2}\langle W_{\mu\nu}W^{\mu\nu}\rangle 
-\frac{1}{4} B_{\mu\nu}B^{\mu\nu}+i\sum_j\bar \psi_j \!\not\!\! D\psi_j+\frac{1}{2}\partial_{\mu}h\partial^{\mu}h\nonumber\\
&+\frac{v^2}{4}\ \langle D_\mu U D^\mu U^{\dagger}\rangle f_U(h)-v\Big[{\bar{\psi}}f_{\psi}(h)UP_{\pm}\psi+{\mathrm{h.c.}}\Big]-V(h)
\end{align}
with
\begin{align}
f_U(h)=1+\sum_ja_j^U\left(\frac{h}{v}\right)^j;\quad f_{\psi}(h)=Y_{\psi}+\sum_jY^{(j)}_{\psi}\left(\frac{h}{v}\right)^j;\quad 
V(h)=\sum_{j\geq 2}a_j^V\left(\frac{h}{v}\right)^j
\end{align}
The absence of arbitrary functions of $h$ next to the Higgs and fermion kinetic terms can be achieved by judicious field redefinitions~\cite{Buchalla:2013rka}. For the gauge kinetic terms this is however not the case and arbitrary powers of $h$ are in principle allowed. We will assume that Higgs couplings to gauge bosons are subleading in the EFT expansion (in a way to be explained below). This is a phenomenological requirement in order to avoid too large contributions, e.g., to $h\to \gamma\gamma$. We will also assume that the new strong sector preserves CP and custodial symmetry in order to comply with experimental bounds. Sources of CP and custodial symmetry breaking are then restricted to the ones already existing in the Standard Model.  
  
From the procedure we followed, it should be obvious that (i) the Standard Model can be recovered from Eq.~(\ref{LO}) for a specific choice of parameters; and (ii) the resulting Lagrangian is in general nonrenormalizable and counterterms will be needed to renormalize it order by order. Related to the previous point, it is evident that an expansion in canonical dimensions is no longer valid and in order to be able to properly define the EFT one first needs to identify the right expansion criteria.  

\begin{figure}[t]
\begin{center}
\includegraphics[width=3.5in]{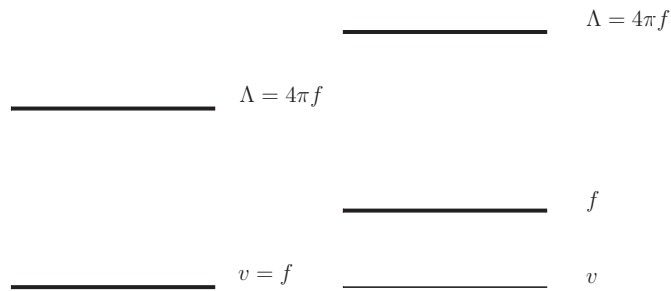}
\end{center}
\caption{\small{\it{Dynamical scales involved in the EFT. The panel in the left illustrates the QCD-like scenario incarnated by technicolor models. The panel in the right corresponds to scenarios with vacuum misalignment, where $v$ and $f$ are generically different.}}}\label{fig:1}
\end{figure}

The generic dynamical picture to have in mind when considering Eq.~(\ref{LO}) is sketched in Fig.~\ref{fig:1}. A generic group is broken spontaneously at a scale $f$ generating a number of Goldstone modes. Since the EFT is minimal, i.e., we are including only the particle content that has been detected, only four Goldstone modes are required. Heavy resonances corresponding to these new (spontaneously broken) interactions appear typically at a cutoff scale $\Lambda\sim 4\pi f$. The scale $v$ is dynamically generated and does not need to coincide with $f$. Weakly-coupled new physics, if present, are assumed to appear at higher scales and therefore have a negligible impact. 

The interplay of the different dynamical scales is therefore described by the dimensionless parameters
\begin{align}
\xi=\frac{v^2}{f^2};\qquad \qquad \ell=\frac{f^2}{\Lambda^2}\sim \frac{1}{16\pi^2}
\end{align}
Since $\Lambda\sim 4\pi f$, the loop expansion is the natural expansion parameter of this theory, while the misalignment parameter $\xi$ is a phenomenological input to be determined experimentally. The present experimental situation sets an upper bound at roughly $\xi\lesssim 0.1$. As I will argue below, this uncertainty is too big to be interpreted as a new-physics effect coming from weakly-coupled dynamics. Notice that $\xi$ is actually the knob that provides a smooth transition between the strongly-coupled regime allowed by the current experimental precision and the Standard Model, which is recovered in the limit $\xi\to 0$.

Before getting into the technicalities of power counting, let me emphasize a nice property of Eq.~(\ref{LO}). Since it represents a modified version of the scalar sector of the Standard Model, deviations in Higgs couplings can be potentially sizable. In contrast, gauge interactions are SM-like and potential deviations should first appear at NLO in the expansion. This generic hierarchy for deviations matches the current experimental status of SM interactions, where the level of scrutiny of LEP for gauge interactions is roughly two orders of magnitude more precise than what LHC has achieved for Higgs interactions. Another remarkable property of Eq.~(\ref{LO}) is that while it naturally allocates room for sizable deviations in the Higgs sector, the EFT is valid even if experimental precision constrains those deviations to smaller values.   

%%%%%%%%%%%%%%%%%%%%%%%%%%%%%%%%%%%%%%%%%%%%%%%%%%%%%%%%%%%%%%%%%%%%%%%%%%%%%

\section{Systematics of the EFT expansion}

\begin{figure}[t]
\begin{center}
\includegraphics[width=6.5cm]{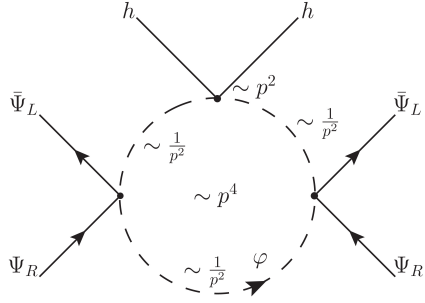}
\end{center}
\caption{\small{\it{Sample diagram illustrating the procedure behind Eq.~(3.1). The superficial divergence of this diagram is absorbed by a local NLO operator of the form $({\bar{\psi}}_L\psi_R)^2 h^2$.}}}
\label{fig:2}
\end{figure}

There are two ways one can proceed in order to define the right power counting for theories like Eq.~(\ref{LO}) and in general for any nonrenormalizable EFT. The hard way is to build the NLO counterterms by absorbing all the one-loop divergences of the theory. The most elegant way of doing so is to build the effective action through the heat kernel method. A more mundane procedure is to examine all the one-loop topologies and extract by induction a formula for their superficial degree of divergence. Since there is only a finite number of interactions this can always be done. A particular example of this latter procedure is provided in Fig.~\ref{fig:2}.  

For the case at hand the structure of any loop diagram in terms of the field content is~\cite{Buchalla:2012qq,Buchalla:2013rka}  
\begin{align}\label{dod}
\Delta=\frac{p^d}{\Lambda^{2L}}\Bigg[(yv)^{\nu_f}\left(\frac{\Psi}{v}\right)^F\Bigg]\Bigg[(gv)^{\nu_g}\left(\frac{X_{\mu\nu}}{v}\right)^G\Bigg]\Bigg[v^2\left(\frac{\varphi}{v}\right)^B\Bigg]\Bigg[(hv)^{2\nu_h}\left(\frac{h}{v}\right)^H\Bigg]
\end{align} 
where 
\begin{align}\label{PC}
d=2L+2-\frac{F}{2}-G-2\nu_h-\nu_f-\nu_g
\end{align}
is the degree of divergence. $F,G,B,H$ count the number of external fields while $\nu_j$ the number of vertices. The first thing to notice is that the degree of divergence is bounded from below. Actually, the presence of gauge fields and fermions reduces the degree of divergence of a diagram. This ensures that the number of counterterms is finite, as one would expect from consistency arguments.  The other thing to notice is that not just fields, but also couplings count. This might seem suprising at first, but is a direct consequence of the loop expansion. Below I will justify why this has to be so.

As a trivial application of Eq.~(\ref{PC}), consider Eq.~(\ref{LO}) just keeping the Goldstone interactions. The result should reproduce the power counting of chiral perturbation theory, and indeed one recovers Weinberg's result, $d=2L+2$~\cite{Weinberg:1978kz}. Weinberg's power-counting formula starts from topological arguments and concludes that the chiral expansion is an expansion in derivatives. In practice this second criterium is what is used in ChPT: one counts the number of derivatives and readily knows the order of an operator. 

This suggests that a similar result could be inferred when gauge bosons, Higgs and fermion interactions are reinstated in Eq.~(\ref{LO}). This brings us to the second way of deriving the power counting. The criterium is to define weights (which we will call chiral dimensions) to the elements in Eq.~(\ref{LO}) such that all the terms are homogeneous, i.e., have the same global weight. Naively one might think that either there is no solution or there are many, but requiring that the result of ChPT is reproduced in the right limit it turns out that there is a unique prescription, namely~\cite{Nyffeler:1999ap,Buchalla:2013eza}
\begin{align}\label{weights}
[\phi; h; X_{\mu}]_{\chi}=0; \qquad [\partial_{\mu}; {\bar{\psi}}\Gamma \psi; g; y]_{\chi}=1
\end{align} 
In hindsight, this result was already imprinted in Eq.~(\ref{PC}). Rearranging terms one can write
\begin{align}
2L+2=d+\frac{F}{2}+G+2\nu_h+\nu_f+\nu_g
\end{align} 
which shows that the (loop) order of the expansion depends on fields, derivatives and couplings in precisely the way dictated by Eq.~(\ref{weights}). Chiral dimensional counting is therefore the weight prescription that reproduces loop counting. This prescription is unambiguous and unique once the leading-order Lagrangian of the theory is specified. From Eq.~(\ref{dod}) one can then build the full set of operators at any order in the loop expansion~\cite{Buchalla:2013eza}. This has been explicitly done at NLO~\cite{Buchalla:2013rka}. Of course, the most difficult task is to ensure that the resulting set of operators is an actual basis and redundancies have been removed.

It should be emphasized that power counting is a way to organize an expansion {\emph{given}} a leading-order Lagrangian. It does not exist as a universal recipe to apply to generic physical systems but rather relies on dynamical insight and the identification of the leading-order operators. For perturbative systems, this comprises the kinetic term for the fields involved plus the dominant interactions. 

As a toy example, consider adding to Eq.~(\ref{LO}) a self-interaction for fermions, such that the fermionic leading-order piece would take the form
\begin{align}\label{ex}
{\cal{L}}&=-i{\bar{f}}\!\not\!\! D f-\lambda ({\bar{f}}\Gamma f)({\bar{f}}\Gamma f)-v {\bar{f}}Y_f(h)UP_{\pm} f
\end{align}
From our previous assignments in~(\ref{weights}), all the terms are homogeneously of chiral dimension two if $[\lambda]_{\chi}=0$. The question is then what prevents fermion bilinears from appearing at leading order in our Lagrangian. The answer is dynamical considerations: the presence of the fermion self-interactions implies that dynamically its mass is close to the cutoff $\Lambda$, i.e., the fermion would not be elementary but a heavy baryon of the strong interactions and, as such, it should be integrated out of the theory. Eq.~(\ref{ex}) is therefore inconsistent. For an elementary fermion, four-fermion operators arise necessarily as counterterms at NLO, i.e., with two powers of weak couplings out front.

%%%%%%%%%%%%%%%%%%%%%%%%%%%%%%%%%%%%%%%%%%%%%%%%%%%%%%%%%%%%%%%%%%%%%%%%%%%%%

\subsection{Chiral dimensions in ChPT}

The power-counting formula that we derived applies generically to nonrenormalizable theories of fermions, gauge fields and scalars which admit a perturbative expansion. Since ChPT falls in this category, it is instructive to reexamine it in the light of Eq.~(\ref{weights}). 

Before delving into details, a general caveat that one should keep in mind. The similarities between QCD and a theory of dynamical EWSB hold to the extent that both theories have their mathematical foundations on nonlinear representations. However, one should bear in mind that in QCD the Goldstone modes arise through breaking of a global $SU(3)_L\times SU(3)_R$ symmetry, while in the electroweak sector the broken symmetry is local. In the former, the Goldstone modes are physical and build up the octet of pseudoscalar mesons ($\pi$'s, $K$'s and $\eta$), while in the latter the Goldstone modes give the $W$ and $Z$ gauge fields a mass. This also implies that there is no analog of the anomalous Wess-Zumino-Witten Lagrangian in the EFT for electroweak interactions, since gauge interactions should be anomaly-free. For power-counting matters, however, the two theories fall into the same category.    
  
Let us consider first $n_f=3$ ChPT coupled to external sources. At leading order the Lagrangian reads~\cite{Gasser:1983yg}
\begin{align}
\mathcal{L}_{LO}&=\frac{f^2}{4}\langle D_{\mu}U^{\dagger}D^{\mu}U\rangle+\frac{f^2}{4}\langle U^{\dagger}\chi+\chi^{\dagger}U\rangle
\end{align}
where $D_{\mu}U=\partial_{\mu}U-i\,r_{\mu}U+i\,U\,l_{\mu}$ and $\chi=2\,B_0\,(s+i\,p)$. As long as external sources are nondynamical their chiral dimension is arbitrary: their sole effect is to source the Green's functions of the theory. In other words, no matter the chiral dimension given to the sources, the contribution to every Green's function will be unaffected.

Things are however different when external sources become dynamical, e.g. by adding pion masses. In that case 
\begin{align}
\chi\sim m_{\pi}^2\sim \frac{\langle {\bar{q}}q\rangle}{f^2}m_q
\end{align}
The usual ChPT prescription~\cite{Gasser:1983yg} correctly sets $m_{\pi}\sim {\cal{O}}(p)$, ensuring that the meson propagator is homogeneous. However, this is traced back to $m_q\sim {\cal{O}}(p^2)$. Since quark masses are $m_q\sim y v$, this would suggest that $y\sim {\cal{O}}(p^2)$. This makes the counting at the quark level cumbersome: the kinetic term for quarks would be inhomogeneous, with the mass term suppressed. Eq.~(\ref{weights}) instead dictates that $[y]_{\chi}=1=[\langle{\bar{q}}q\rangle]_{\chi}$, while the breaking scales $v$ and $f$ have no chiral dimension. Notice that this consistently brings $[m_{\pi}]_{\chi}=1$ while keeping the homogeneity of the Lagrangian both at the pion and the quark level. 

As a second example, consider ChPT with dynamical photons. Now the leading order is~\cite{Urech:1994hd} 
\begin{align}\label{em}
\mathcal{L}_{LO}&=\frac{f^2}{4}\langle D_{\mu}U^{\dagger}D^{\mu}U\rangle+\frac{f^2}{4}\langle U^{\dagger}\chi+\chi^{\dagger}U\rangle-\frac{1}{4}F_{\mu\nu}F^{\mu\nu}+e^2\delta\langle U^{\dagger}QUQ\rangle
\end{align}
where $D_{\mu}U=\partial_{\mu}U+ieA_{\mu}[Q,U]$ and $Q=$diag$(2/3,-1/3,-1/3)$. The last operator is induced by virtual photons and gives a contribution to the pion potential, providing in particular electromagnetic corrections to pion masses proportional to 
\begin{align}
(m_{\pi}^2)_{{\mathrm{em}}}\sim e^2\frac{\delta}{f^2}
\end{align}
Since $\delta$ is loop-induced, all the terms in Eq.~(\ref{em}) are actually of the same order, i.e., their characteristic energy scale is $f$. When QED is switched on, it is usually argued that the chiral expansion gets upgraded to a double expansion in $e^2$ and $p$. While this is technically correct, it obscures the systematics of the chiral expansion. In particular, it is not clear what to do if the electric charge were numerically bigger. 

Eq.~(\ref{weights}) adapted to the present case shows that a double expansion is actually not required: consistency of the theory is achieved with a single expansion, with chiral weights given by
\begin{align}
[A_{\mu}]_{\chi}=0;\qquad [e]_{\chi}=1;\qquad [\partial_{\mu}]=1
\end{align}
Notice that with this prescription all the operators in Eq.~(\ref{em}) are homogeneously of chiral dimension two. Using Eq.~(\ref{dod}) adapted to the present case the full set of NLO counterterms can be generated, regardless of the size of the electric charge. This shows that the structure of divergences and counterterms of the theory is independent of the actual size of $e$.\footnote{Of course, this makes sense as long as the coupling is still perturbative, but the point is that a much larger value of $e$ would not alter the structure and systematics of the EFT.} Obviously, one can always use the fact that $e$ is numerically small to neglect certain counterterms and further simplify the expansion. For instance, the following sample of operators 
\begin{align}
\langle D_{\mu}U^{\dagger}D^{\mu}U\rangle^2,\qquad e^2\langle D_{\mu}U^{\dagger}D^{\mu}U\rangle\langle U^{\dagger}QUQ\rangle\nonumber\\
\langle U^{\dagger}\chi+\chi^{\dagger}U\rangle^2,\qquad e^4\langle U^{\dagger}QUQ\rangle^2
\end{align}
have all chiral dimension four but have a different numerical impact on physical observables. 

While at the QCD scale this systematic aspect might appear superfluous, at the electroweak scale it is of fundamental importance. There the gauge couplings are not suppressed and the formalism of chiral dimensions provides the right description of the physics.

A last comment concerns the chiral dimensions for the gauge sector. Consider ChPT with external (nondynamical) sources at NLO~\cite{Gasser:1983yg}:
\begin{eqnarray}\label{chiralexp4}
\mathcal{L}_4 & = & L_1\,\langle
D_{\mu}{U}^{\dagger}D^{\mu}U\rangle^2+L_2\,\langle
D_{\mu}{U}^{\dagger}D_{\nu}U\rangle\,\langle
D^{\mu}{U}^{\dagger}D^{\nu}U\rangle+\nonumber\\ & + &
L_3\,\langle D_{\mu}{U}^{\dagger}D^{\mu}U\,
D_{\nu}{U}^{\dagger}D^{\nu}U\rangle+L_4\,\langle
D_{\mu}{U}^{\dagger}D^{\mu}U\rangle\,\langle{U}^{\dagger}\chi+{\chi}^{\dagger}U\rangle+\nonumber\\
& + & L_5\,\langle
D_{\mu}{U}^{\dagger}D^{\mu}U\,({U}^{\dagger}\chi+{\chi}^{\dagger}U)\rangle+L_6\,\langle\,{U}^{\dagger}\chi+{\chi}^{\dagger}U\rangle^2+\nonumber\\
& + &
L_7\,\langle\,{U}^{\dagger}\chi-{\chi}^{\dagger}U\rangle^2+L_8\,\langle\,{\chi}^{\dagger}U{\chi}^{\dagger}U+{U}^{\dagger}\chi{U}^{\dagger}\chi\rangle-\nonumber\\
& - & i\,L_9\,\langle
F_R^{\mu\nu}D_{\mu}UD_{\nu}{U}^{\dagger}+F_L^{\mu\nu}D_{\mu}{U}^{\dagger}D_{\nu}U\rangle+L_{10}\,\langle\,{U}^{\dagger}F_R^{\mu\nu}UF_{L\mu\nu}\rangle+\nonumber\\
  & + & H_1\,\langle
  F_{R\mu\nu}F_R^{\mu\nu}+F_{L\mu\nu}F_L^{\mu\nu}\rangle+H_2\,\langle{\chi}^{\dagger}\chi\rangle
\end{eqnarray}
with
\begin{equation}
F_L^{\mu\nu}=\partial^{\mu}l^{\nu}-\partial^{\nu}l^{\mu}-i\,[l^{\mu},l^{\nu}];\qquad
F_R^{\mu\nu}=\partial^{\mu}r^{\nu}-\partial^{\nu}r^{\mu}-i\,[r^{\mu},r^{\nu}]
\end{equation} 
The external fields $l_{\mu}, r_{\mu}$ are typically counted as ${\cal{O}}(p)$. One could naively argue that if gauge fields do not carry chiral dimensions, as we argued above, the operators proportional to $L_{10}$ and $H_1$ would appear with chiral dimension two, while $L_9$ would count as chiral dimension three. This is not the case. When external gauge fields become physical the gauge coupling has to be specified. For instance, the operator with $L_{10}$ becomes, when a dynamical photon is included,     
\begin{align}
e^2F_{\mu\nu}F^{\mu\nu}\langle U^{\dagger}QUQ\rangle
\end{align} 
which is clearly of chiral dimension four. The same applies to $L_9$ and $H_1$.\footnote{$H_1$ is actually slightly more subtle: with dynamical fields it takes the form of a kinetic term with two powers of the gauge coupling. This term becomes redundant, since it can be reabsorbed once the photon kinetic term is brought to canonical form.} 

Maybe the cleanest example that chiral dimensions are the correct counting in ChPT is provided when dynamical photons and leptons are also included. The leading order Lagrangian then reads
\begin{align}\label{emf}
\mathcal{L}_{LO}&=\frac{f^2}{4}\langle D_{\mu}U^{\dagger}D^{\mu}U\rangle+\frac{f^2}{4}\langle U^{\dagger}\chi+\chi^{\dagger}U\rangle-\frac{1}{4}F_{\mu\nu}F^{\mu\nu}+e^2\delta\langle U^{\dagger}QUQ\rangle+i\sum_j\bar \ell_j (\!\not\!\! D-m_{\ell})\ell_j
\end{align}
The full set of NLO operators has been worked out using the heat kernel method~\cite{Knecht:1999ag}. With this method one does not need to specify an explicit power counting: the result already contains the set of operators that absorb the one-loop divergences. One can explicitly check that the resulting list of counterterms given in~\cite{Knecht:1999ag} are consistently of chiral dimension four.

The previous examples show that chiral dimensions are a unique and consistent prescription for nonrenormalizable EFT expansions. The fundamental difference between linear {\it{vs}} nonlinear expansions can be seen in the fact that chiral dimensions cannot be reduced to a counting in canonical dimensions. A dramatic example is provided by the fermionic operators
\begin{align}
e^2({\bar{\ell}}_i\gamma_{\mu}\ell_i)({\bar{\ell}}_j\gamma^{\mu}\ell_j),\qquad 
e^2({\bar{\ell}}_i\gamma_{\mu}\ell_i)\langle D^{\mu}U^{\dagger}QUQ\rangle
\end{align}
which show up at NLO in ChPT coupled to QED but clearly have different canonical dimensions.  

%%%%%%%%%%%%%%%%%%%%%%%%%%%%%%%%%%%%%%%%%%%%%%%%%%%%%%%%%%%%%%%%%%%%%%%%%%%%%

\section{The vacuum misalignment mechanism}

In the previous sections I have insisted that the nonlinear EFT is organized in loops, while the decoupling parameter $\xi$ is basically a knob that can be determined from fitting to experimental data. Since $\xi$ is not associated with the power counting, it must be implicitly contained inside the Wilson coefficients of the EFT. 

The simplest example to see these ideas at play is the model first developed in~\cite{Agashe:2004rs}. One considers the group $G=SO(5)\times U(1)_X$ broken to $H=SO(4)\times U(1)_X$ at a scale $f$. The latter is isomorphic to $SU(2)_L\times SU(2)_R\times U(1)_X$ and the Standard Model can be recovered through the usual pattern of electroweak symmetry breaking at the scale $v$. Here I will concentrate on the bosonic sector of the theory. The fermionic sector is important to build a realistic Higgs potential but it is also more model-dependent and can be skipped if one is primarily interested in discussing the interplay of the dynamical scales. 

From the pattern of symmetry breaking at $f$ one generates 4 Goldstone bosons $h^A$, which build a vector in the fundamental representation of $SO(4)$. They are usually parametrized as
\begin{align}
\Sigma(h^A)={\cal{U}}(x)\Sigma_0,\qquad \Sigma_0=\left(\begin{array}{c}
0_4\\1
\end{array}\right)\nonumber
\end{align}   
where ${\cal{U}}={\mathrm{exp}}(\sqrt{2}it^Ah^A/f)$ and $t^A$ are the broken generators.

\begin{figure}[t]
\begin{center}
\includegraphics[width=6.5cm]{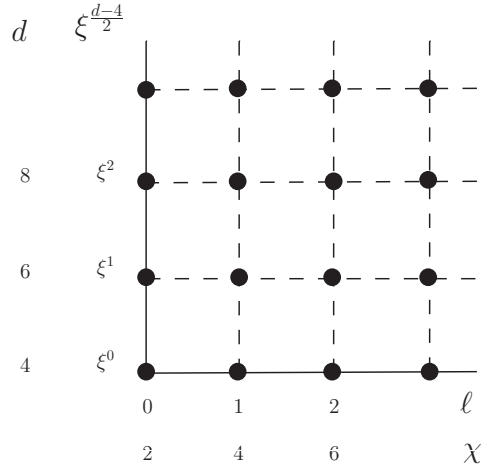}
\end{center}
\caption{\small{\it{Graphical illustration of the double expansion in loops and the decoupling parameter $\xi$. As argued in the main text, the nonlinear Lagrangian amounts to a resummation of the vertical lines order by order in the loop expansion, which is the parameter controlling the expansion. The $\xi$ expansion is numerical (not linked to power counting) and fixed by current experimental data.}}}\label{fig:3}
\end{figure}

In order to bring the notation to the one used in the construction of the EFT, one can use the isomorphism between $SO(4)$ and $SU(2)\times SU(2)$ to express the previous equation in terms of the $SU(2)$ bidoublet matrix $U$ and a radial Higgs. The isomorphism reads $H=h_A\lambda_A\equiv h U$ with $\vec{\lambda}=(i{\vec{\sigma}},1_2)$. Thus it follows that
\begin{align}
h_A=\frac{h}{2}\langle U\lambda_A^{\dagger}\rangle\nonumber
\end{align}
and one can express $\Sigma$ as
\begin{align}
\Sigma(h,U)=\left(\begin{array}{c}
\displaystyle\frac{\langle U\lambda_A^{\dagger}\rangle}{2}\sin{h/f}
\\ \cos{h/f}
\end{array}\right)\nonumber
\end{align}
This is the building block from which to build bosonic operators. At leading order one finds
\begin{align}\label{SO5}
{\cal{L}}_{LO}=\frac{f^2}{2}\langle D_{\mu}\Sigma^{\dagger}D^{\mu}\Sigma\rangle-V&=\frac{1}{2}\partial_{\mu}h\partial^{\mu}h+\frac{f^2}{4}\langle D_{\mu}U^{\dagger}D^{\mu}U\rangle\sin^2\frac{h}{f}-V
\end{align}
where the potential is built from the two $SO(5)$-breaking spurions (see~\cite{Buchalla:2014eca} for details) and eventually takes the form~\cite{Contino:2010rs} 
\begin{align}
V=\alpha \cos{h/f}-\beta \sin^2{h/f}
\end{align} 
with $\alpha$ and $\beta$ such that $\beta>0$ and $|\alpha|\leq 2\beta$ (see also~\cite{Contino:2010rs} for a concrete realization of the potential). Under such circumstances there is spontaneous symmetry breaking and the Higgs acquires a nontrivial vacuum expectation value. Expanding around the vacuum, $h=\langle h \rangle+{\hat{h}}$, Eq.~(\ref{SO5}) should match our EFT parametrization, i.e., 
\begin{align}
{\cal{L}}_{LO}&=\frac{1}{2}\partial_{\mu}{\hat{h}}\partial^{\mu}{\hat{h}}+\frac{v^2}{4}\langle D_{\mu}U^{\dagger}D^{\mu}U\rangle f_U({\hat{h}})-V({\hat{h}})
\end{align}
In particular, going to unitary gauge and matching to the gauge boson masses one finds that $v$ is given by
\begin{align}
v=f\sin\frac{\langle h \rangle}{f}\nonumber
\end{align}
and therefore the decoupling parameter in this model takes the simple form $\xi=\sin^2{\langle h \rangle/f}$. In terms of $\xi$, $f_U$ and $V$ can then be expressed as
\begin{align}
f_U({\hat{h}},\xi)&=\displaystyle\cos\frac{2{\hat{h}}}{f}+\sqrt{\frac{1-\xi}{\xi}}\sin\frac{2\hat{h}}{f}+\frac{1}{\xi}\sin^2\frac{\hat{h}}{f}=1+2\sqrt{1-\xi}\displaystyle\left(\frac{\hat{h}}{v}\right)+(1-2\xi)\left(\frac{\hat{h}}{v}\right)^2+{\cal{O}}\left({\hat{h}}^3\right)
\end{align}
and
\begin{align}
V({\hat{h}},\xi)=\frac{v^2}{2}m_h^2 \left[\left(\frac{{\hat{h}}}{v}\right)^2+\sqrt{1-\xi}\left(\frac{{\hat{h}}}{v}\right)^3+\frac{3-7\xi}{12}\left(\frac{{\hat{h}}}{v}\right)^4+{\cal{O}}\left({\hat{h}}^5\right)\right]
\end{align}
In a specific model like the one discussed here the dependence of the Wilson coefficients on $\xi$ can be fully determined. In the EFT the $\xi$ dependence is implicit. However, experimentally $\xi$ is small, so one can expand in it. In this scenario the EFT can be seen as a double expansion: the loop expansion associated with the nonrenormalizability of the theory and the numerical expansion in $\xi$. The situation is depicted in Fig.~\ref{fig:3}, where each point is identified by its coordinates $(\ell, \xi)$. Vertical lines in the chart correspond to each order in the loop expansion. The Standard Model corresponds to the dot in the bottom left corner, i.e., LO both in the loop and $\xi$ expansions, while SM loops are matched to the horizontal axis. 

Correspondingly, notice that the operators belonging to the point $(0,1)$ are $\xi$-suppressed with respect to the Standard Model. Therefore, the $\xi$ expansion functions like an expansion in canonical dimensions (in inverse powers of $f^2$). As a result the operators of $(0,1)$ can be formally matched to (a subset of) operators from the basis of canonical dimension six.\footnote{The rest of the dimension-6 operator basis belongs to the point $(1,1)$. Dimension-6 operators are split in two points because some of them bear a loop suppression, a piece of information that the power-counting still provides.}

In general, the identification of ${\cal{O}}(\xi^n)$ contributions in the nonlinear EFT can be done by reexpressing the basis of canonical dimension $2n+4$ in polar coordinates using          
\begin{align}
\sqrt{2}(\tilde{H},H)=(v+h)U
\end{align}
The computation of ${\cal{O}}(\xi)$ was done in~\cite{Buchalla:2014eca}. Technical details about the matching procedure can be found there. 

Despite the previous formal identification of operators in the matching procedure described above and the fact that the expansion in $\xi$ is a canonical one, the power counting is still the one of a nondecoupling, nonrenormalizable EFT: one can reproduce the Standard Model by decoupling the strong new physics, but as long as $\xi$ is nonzero the dynamics do not correspond to a weakly-coupled extension of the Standard Model.
  
%%%%%%%%%%%%%%%%%%%%%%%%%%%%%%%%%%%%%%%%%%%%%%%%%%%%%%%%%%%%%%%%%%%%%%%%%%%%%

\section{EFT-based fit to LHC Higgs data}

From a phenomenological point of view, the EW nonlinear EFT has two main virtues: (i) it is a generalization of the Standard Model that allows to test the Higgs hypothesis; and (ii) it naturally implements a hierarchy between Higgs and gauge couplings, such that it can easily accommodate departures in the former while keeping the latter in agreement with LEP bounds. 

It is important to stress that both points are absent in a linear EFT: in an expansion in canonical dimensions the SM Higgs is assumed to be correct and therefore deviations are suppressed by powers of $1/\Lambda^2$ both in the Higgs and gauge sectors. While the Higgs might well turn out to be the SM one, from an experimental point of view this assumption is at present unjustified: the current experimental precision on Higgs couplings is around $10-15\%$ at the most and even after Run 3 at the LHC the precision will hardly reach ${\cal{O}}(5\%)$.     

At Run 1, Higgs couplings have been tested using the so-called $\kappa$ formalism, a signal-strength parametrization
at the level of the decay rates and production cross sections. The $\kappa$ formalism has been sometimes criticised as being an {\it{ad hoc}} phenomenological parametrization not based on QFT principles. However, until recently it has not been appreciated that it is basically equivalent to the EW nonlinear EFT at leading order~\cite{Buchalla:2015wfa}. In order to make this connection transparent nonlocal parameters have to be traded by local ones.
 
Taking into account the processes so far tested at the LHC, the nonlinear EFT in the unitary gauge leads to the following parametrization: 
\begin{align}\label{fit}
{\cal{L}}&=2c_V\left(m_W^2W_{\mu}W^{\mu}+\frac{1}{2}Z_{\mu}Z^{\mu}\right)\frac{h}{v}-\sum_{f=t,b,\tau}c_fy_f{\bar{f}}fh+c_{gg}\frac{g_s^2}{16\pi^2}G_{\mu\nu}G^{\mu\nu}\frac{h}{v}+c_{\gamma\gamma}\frac{e^2}{16\pi^2}F_{\mu\nu}F^{\mu\nu}\frac{h}{v}
\end{align}
in terms of just 6 parameters. The previous Lagrangian can be easily extended to cover processes like $h\to Z\gamma$, $h\to \mu\mu$, associated production or double Higgs production, to be tested in Run 2.
\begin{figure}[t]
\begin{center}
\vskip -0.7cm
\includegraphics[width=7.0cm]{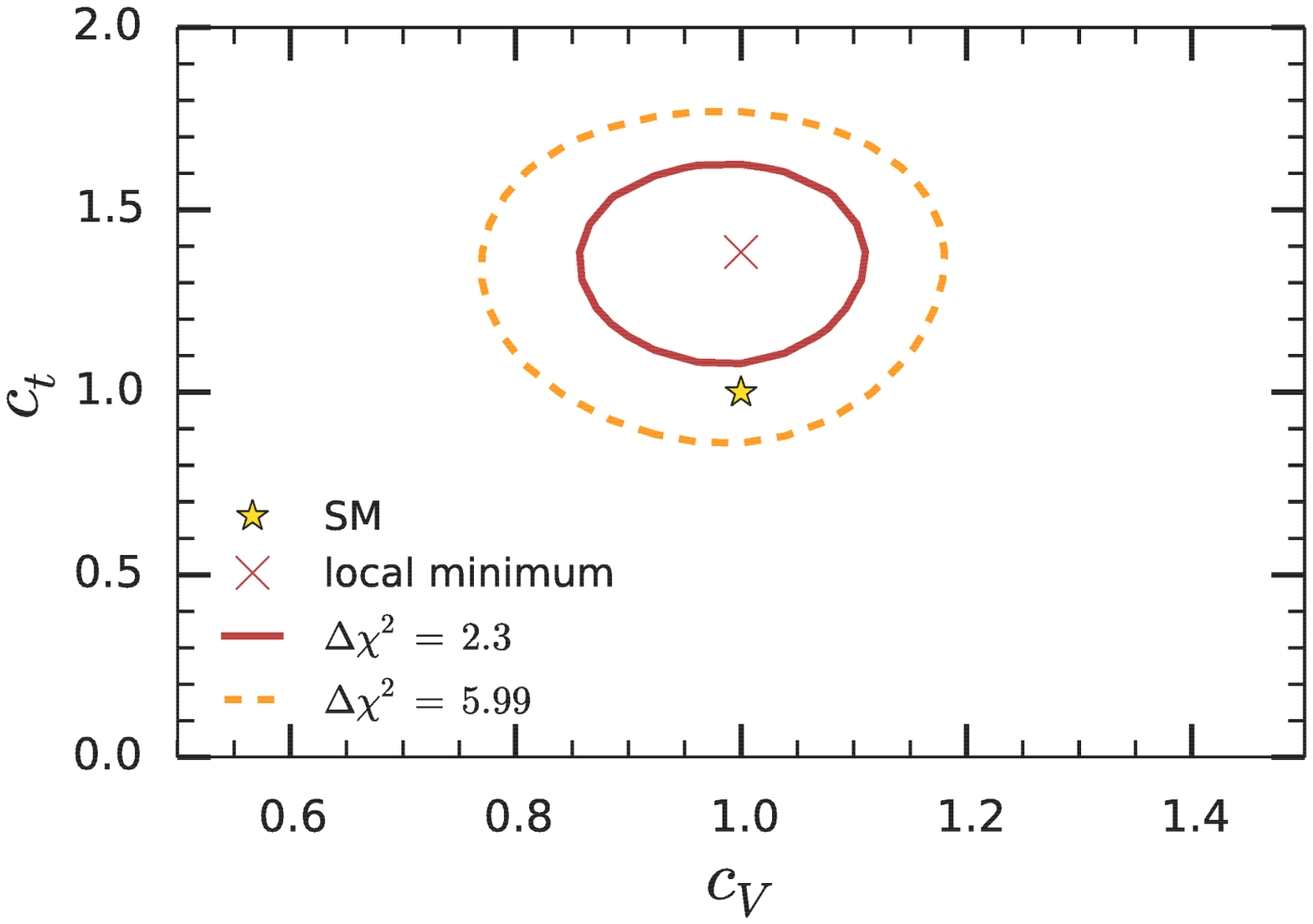}
\includegraphics[width=7.0cm]{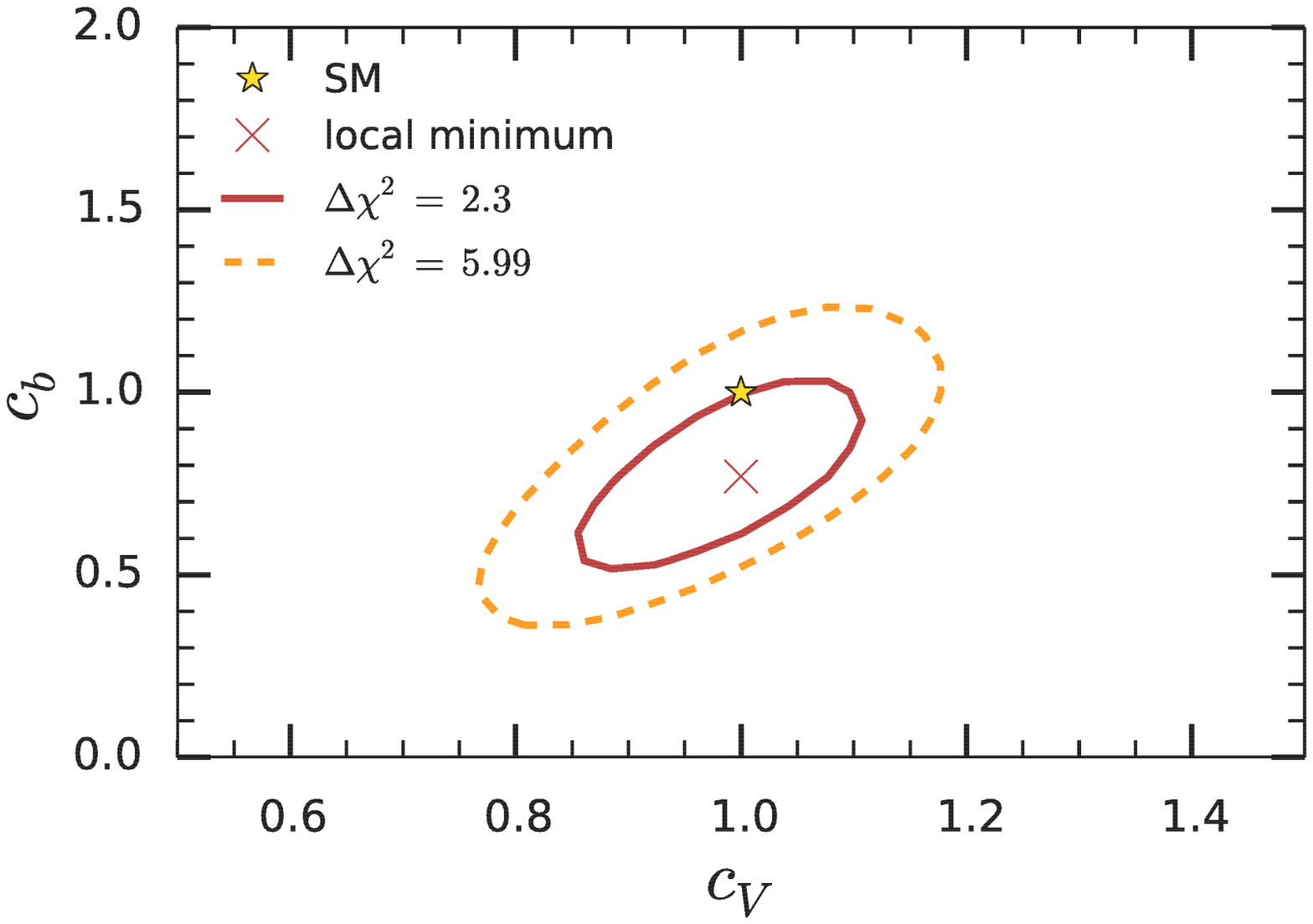}
\vskip -0.0cm
\includegraphics[width=7.1cm]{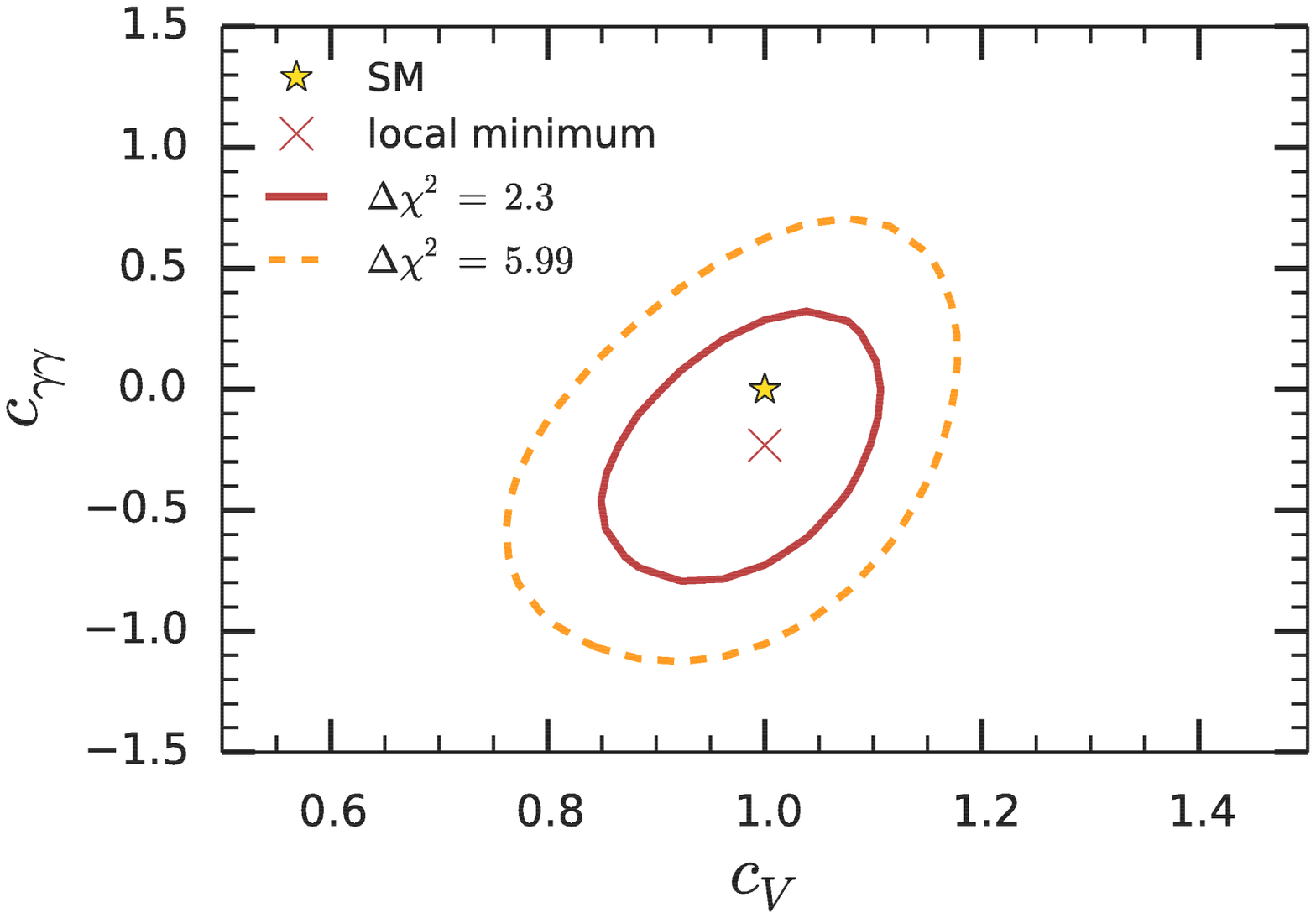}
\includegraphics[width=7.1cm]{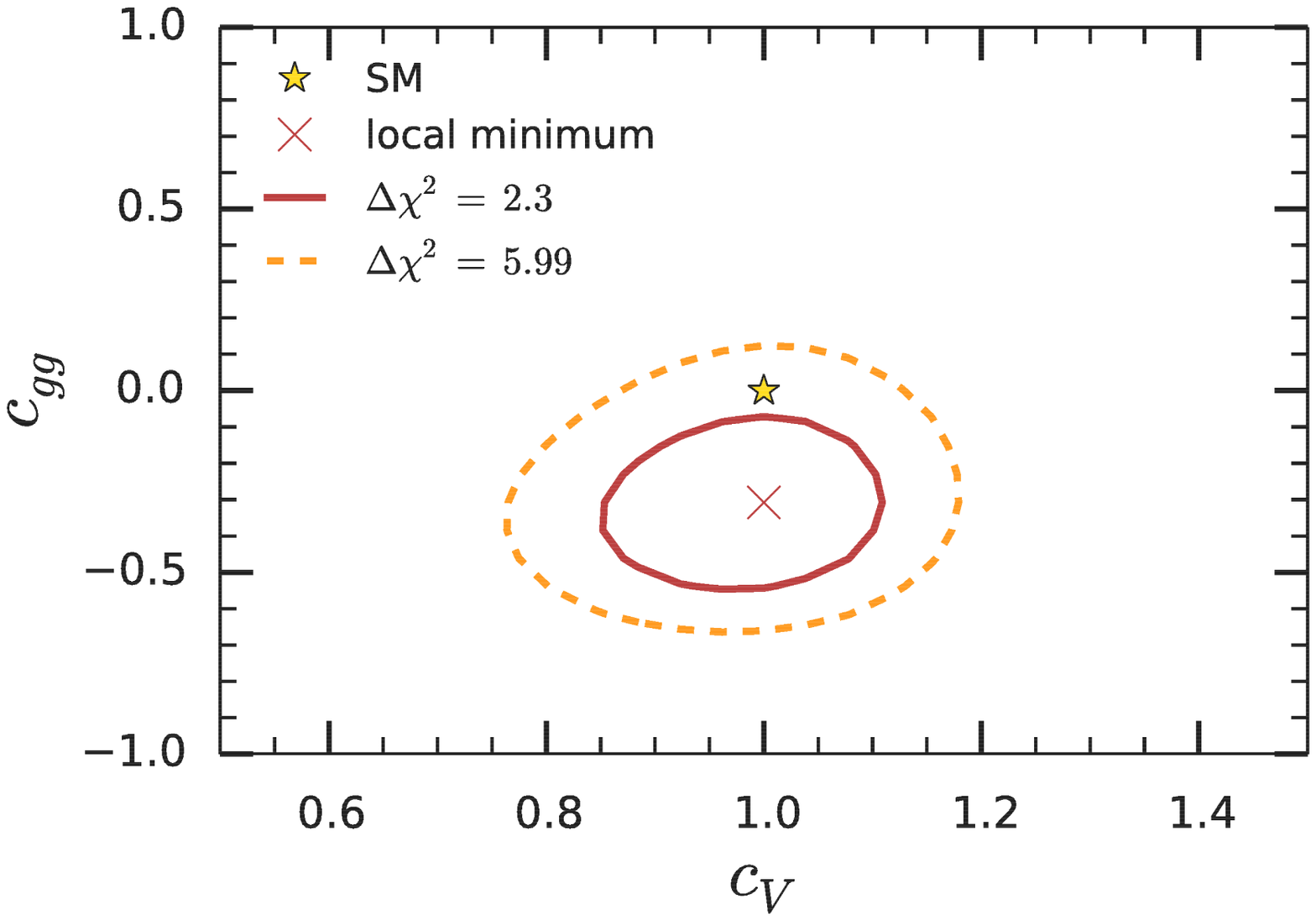}
\vskip -0.0cm
\includegraphics[width=7.0cm]{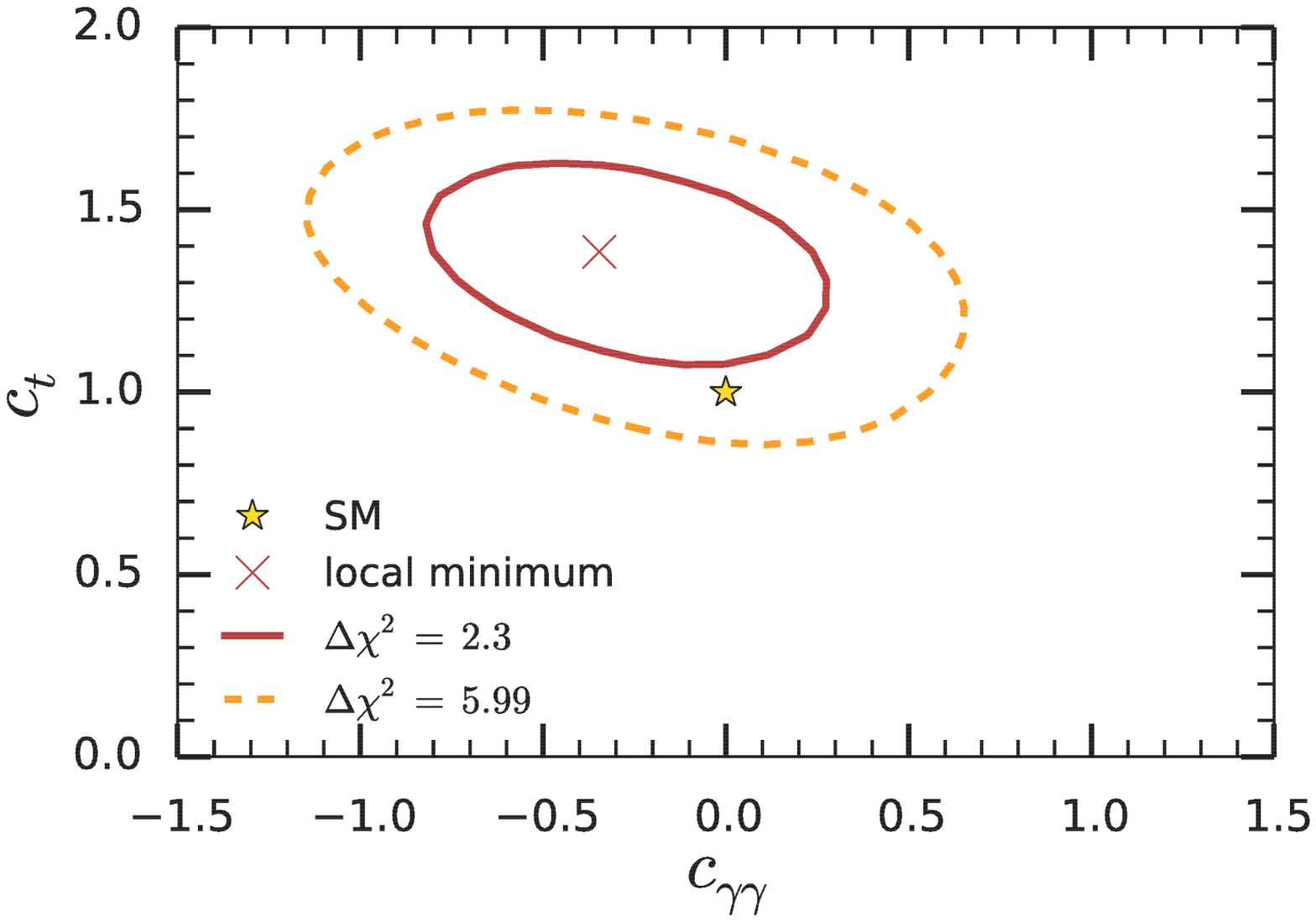}
\includegraphics[width=7.0cm]{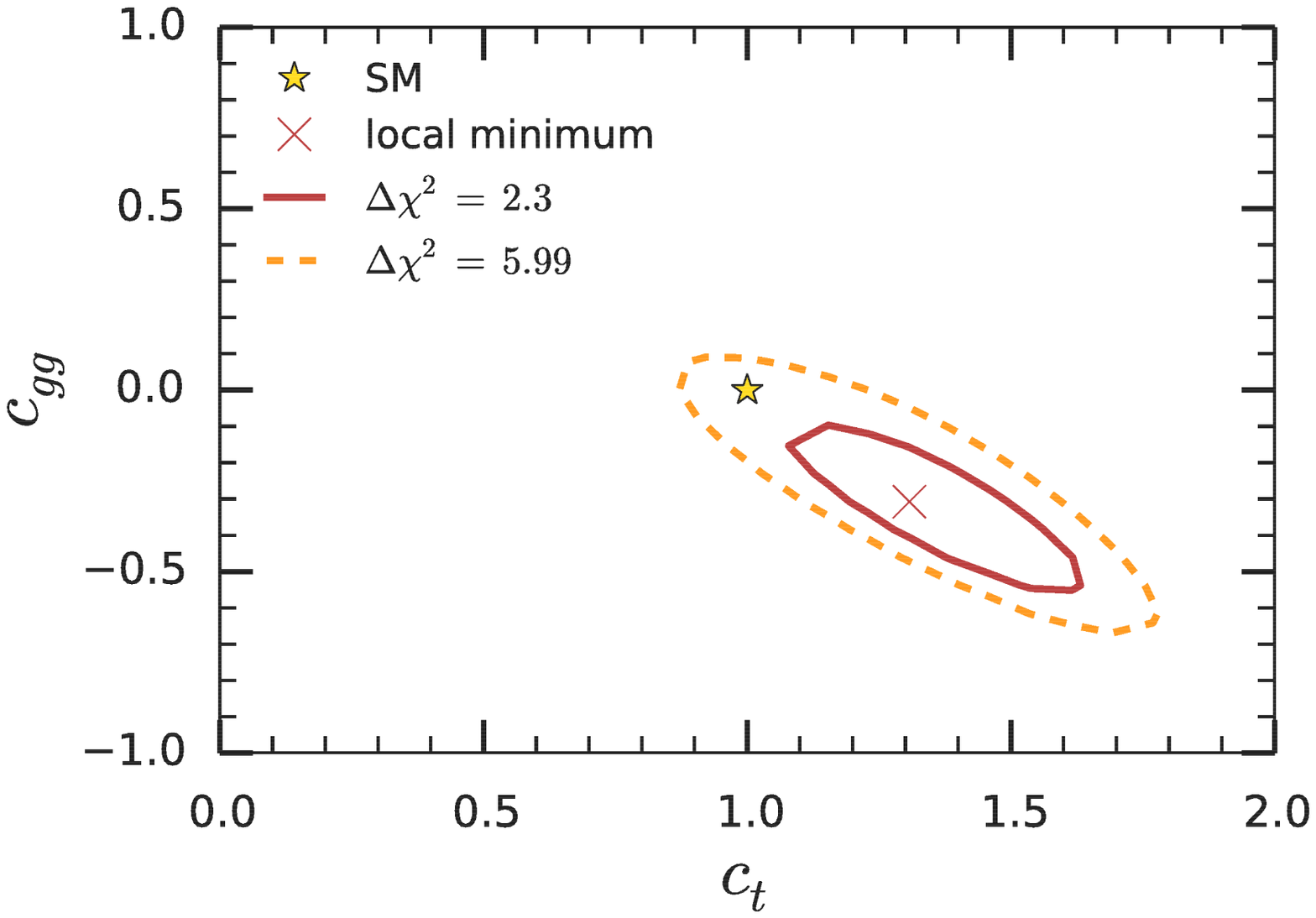}
\end{center}
\caption{\small{\it{$\Delta \chi^2$ isocontours for a sample of two-dimensional plots with the parameters given in Eq.~(5.1)~\cite{Buchalla:2015qju}. In all cases, marginalization over the remaining four parameters has been performed.}}}\label{fig:4}
\end{figure}

The advantages of having an EFT framework behind a phenomenological fitting scheme are multiple. First, it guarantees that QFT basic rules are respected, e.g. gauge invariance. Second, the scheme can be systematically improved as experimental precision demands it. For instance, it is easy to incorporate NLO effects to study shapes in particular Higgs processes~\cite{Buchalla:2015qju}. Finally, it allows to have a clear picture of the expected sizes of the coefficients. For instance, for $h\to Zl^+l^-$ power-counting considerations imply that deviations in the shapes are suppressed typically by two orders of magnitude $(\sim 1/16\pi^2)$ with respect to deviations in the rate~\cite{Buchalla:2013mpa}. In the linear EFT both would be at the same order, with deviations of typically a few percent. 

The dynamical information contained in the power counting can be incorporated as priors in Bayesian fits to experimental data. An example of such a fit has been performed in~\cite{Buchalla:2015qju}. A sample of two-dimensional projections is shown in Fig.\ref{fig:4}. The results confirm that the current precision in Higgs couplings is at the $10-15\%$ level, with deviations from the Standard Model within 1-2 $\sigma$.  

%%%%%%%%%%%%%%%%%%%%%%%%%%%%%%%%%%%%%%%%%%%%%%%%%%%%%%%%%%%%%%%%%%%%%%%%%%%%%

\section{Conclusions}

Scenarios of dynamical EWSB are still viable candidates for physics beyond the Standard Model. The discovery of the Higgs excludes the simplest models but there are extensions that accommodate a light Higgs-like scalar in a natural way, i.e. through symmetry arguments. Particularly interesting are dynamical settings with vacuum misalignment, where the Higgs is a pseudo-Goldstone boson. In these settings the new strong dynamics can be decoupled, and accordingly there is a smooth limit where the Standard Model can be recovered.

These dynamical ideas can be made model-independent by embedding them into an EFT formalism. The advantage of such an EFT is that it generalizes the Standard Model and therefore provides a formalism to test the Higgs hypothesis. In contrast to a linear EFT of the Standard Model, deviations in the Higgs sector can be sizable while keeping gauge interactions well within LEP bounds. This pattern is a natural consequence of having a strong sector and is naturally implemented in the EFT. Given the current (and prospective) uncertainty in Higgs couplings at the LHC, such an EFT provides an excellent tool for phenomenological studies. More so since one can show that the so-called $\kappa$ formalism currently used by ATLAS and CMS for testing Higgs couplings can be embedded within this EFT, thereby acquiring a full-fledged QFT basis.   

In this write-up I have emphasized the systematics underlying the construction of such an EFT. In particular, I have stressed the similarities between this nonlinear EFT and ChPT with dynamical gauge fields and fermions, especially from the systematic point of view. EFTs are expansions in terms of $\mu/\Lambda$, with $\mu$ a typical energy scale and $\Lambda$ the cutoff of the theory. In nonrenormalizable EFTs the organizing principle of the expansion (the power counting) is based on a loop expansion, where NLO counterterms absorb the divergences generated at leading order. Careful scrutiny shows that the loop expansion is equivalent to an expansion in chiral dimensions, understood as weights on fields and couplings. This correspondence is unique and allows to revisit the ChPT systematics from a new, more general, perspective.     

I have not touched upon phenomenological studies of specific processes, where the nonlinear EFT can give signatures which differ substantially from the ones of the linear EFT. As sample representatives, one could mention an analysis of $h\to Zl^+l^-$~\cite{Buchalla:2013mpa} and some studies oriented mainly to gauge boson production at linear colliders~\cite{Buchalla:2013wpa}. 

As a closing remark, it is interesting to point out~\cite{Cata:2015lta} that also in the EFTs for flavor physics at hadronic scales, despite the fact that the Higgs boson does not appear explicitly, different patterns among the Wilson coefficients can discern strong from weak EWSB. Therefore, even in flavor factories there is potential to disentangle the nature of the Higgs boson.

%%%%%%%%%%%%%%%%%%%%%%%%%%%%%%%%%%%%%%%%%%%%%%%%%%%%%%%%%%%%%%%%%%%%%%%%%%%%%

\end{document}